\documentclass[12pt,a4paper]{article}

\usepackage{amssymb}
\usepackage{amsmath}
\usepackage{amsfonts}
\usepackage{graphicx, float, tabularx}
\usepackage{color}
\usepackage{enumerate}
\usepackage[american]{babel}
\usepackage{soul}
\usepackage{cancel}

\usepackage{hyperref}
\hypersetup{
	pdftitle={Geometric model of black hole quantum $N$-portrait, 
extradimensions and thermodynamics},
	bookmarks=true,
	colorlinks,
    linkcolor={red!50!black},
    citecolor={blue!50!black},
    urlcolor={blue!80!black}
}

\usepackage[draft,colorinlistoftodos,shadow]{todonotes}
\presetkeys{todonotes}{inline}{}

\newcommand{\Mpl}{M_{\rm P}}
\newcommand{\ellp}{L_{\rm P}}

\renewcommand{\d}{\mathrm{d}}
\newcommand{\dd}[2]{\frac{\mathrm{d} #1}{\mathrm{d} #2}}

\newcommand{\C}[1]{ {\cal #1}}
\newcommand{\diag}{ \operatorname{diag}}

\usepackage[nosort]{cite}
\usepackage{color,colordvi}

\begin{document}
\begin{center}
{\Large\bf  Geometric model of black hole quantum $N$-portrait, 
extradimensions and thermodynamics}

\end{center}

\vspace{-0.1cm}

\begin{center}

{\bf 
Antonia M. Frassino}
$^{a,b}$\footnote{frassino@fias.uni-frankfurt.de},  {\bf 
Sven K\"{o}ppel
}
$^{a,b}$\footnote{koeppel@fias.uni-frankfurt.de}
{\bf 
and 
Piero Nicolini}
$^{a,b}$\footnote{nicolini@fias.uni-frankfurt.de}

\vspace{.6truecm}

{\em $^a$Frankfurt Institute for Advandced Studies (FIAS)\\
Ruth-Moufang-Str. 1, 60438 Frankfurt am Main, Germany}\\

{\em $^b$Institut f\"ur Theoretische Physik,\\
Johann Wolfgang Goethe-Universit\"at Frankfurt am Main\\Max-von-Laue-Str. 1
60438 Frankfurt am Main, Germany}\\

\end{center}

\vspace{0.1cm}

\begin{abstract}
\noindent  
 {
\small
Recently a short scale modified black hole metric, known as holographic metric, has been proposed in order to capture the self-complete character of gravity. In this paper we show that such a metric can  reproduce some geometric features expected from the quantum $N$-portrait beyond the semi-classical limit.  We show that for a generic $N$ this corresponds to having an effective energy momentum tensor in Einstein equations or, equivalently, non-local terms in the gravity action. We also consider the higher dimensional extension of the metric and the case of an AdS cosmological term. We provide a detailed thermodynamic analysis of both cases, with particular reference to the repercussions on the Hawking-Page phase transition. }

\end{abstract}

\thispagestyle{empty}
\clearpage
\section{Introduction}

Black holes are among the most fascinating and mysterious objects in Physics and are subject to an intense theoretical and observational research activity. Astrophysical black holes, whose masses typically range from few solar masses to billions of solar masses, are pretty well described by general relativity. There might exist, however, a new breed of black holes with masses below $10^{12}\ \mathrm{kg}$, that have been produced in early Universe epochs due to the extreme matter density fluctuations \cite{CaH74} and/or for quantum mechanical decay of deSitter space \cite{MaR95,BoH96,MaN11}.  Such black holes, one can term microscopic black holes, have sizes of elementary particles and are subject to quantum mechanical effects. 
Even if the existence of microscopic black holes still needs corroboration from experiments and observations, they are extremely interesting since they represent one of the crucial challenges to our understanding of fundamental physics. In the mid 1970's, Hawking showed that quantum effects allow black holes for evaporating, \textit{i.e.}, emitting a thermal spectrum of particles like a black body \cite{Haw75}. 
This fact has supported the idea of black hole thermodynamics. Black holes have a temperature proportional to their surface gravity and an entropy proportional to their horizon area. Such an intriguing scenario is, however, not free of problems. Static, spherically symmetric black holes have a negative heat capacity. This is equivalent to saying that, while emitting particles, they shrink and become hotter and hotter. The process is runaway and the destiny of an evaporating black hole is, in general, still unknown. Another issue concerns the statistical interpretation of the black hole entropy. We still miss a complete understanding of the microscopic degrees of freedom that support the entropy/area law in the thermodynamic limit. A related problem concerns the information loss. This arises already at classical level. The collapsing star microstates become inaccessible when the event horizon forms. The Hawking radiation, however, aggravates the situation. The initial stellar particle pure state would evolve into a mixed state of thermally distributed particles, in marked contrast with one of the principles of quantum mechanics (see \cite{Haw14,Haw15,HPS16,Dva15,SaS15,ADG16} for recent discussions on the topic). 

It is believed that the open problems listed above can only be solved by the ultraviolet completion of the gravitational field. In other words, at short scales, \textit{i.e}, lengths of the order of the Planck length, the spacetime description in terms of general relativity breaks down and a quantum theory of gravity must be invoked. 
This explains the great interest in new black hole models able to account for quantum gravity effects in place of the classical curvature singularity (see \textit{e.g.} \cite{BoR00,NSS06a,NSS06b,Riz06,ANS07,SSN09,Nic09,NiS10,SmS10,
MoN10b,Mod06,MMN11,Nic12,IMN13,GrS08,SaS14}).  It is interesting to note that these models tend to converge towards a unique scenario: the singularity is removed or softened; the evaporating black hole undergoes a phase transition prior the final stages of evaporation, switching from a negative heat capacity warming to a positive heat capacity cooling down. 
Ultimately the black hole approaches a zero temperature stable black hole  remnant configuration. Alternatives to the above scenario are offered by the Planck star \cite{DPR15} and by some models inspired by the generalized uncertainty principle  that allow for either the formation of hot remnants \cite{APS01} or of sub-Planckian black holes \cite{CMN15} (for recent reviews see \cite{DeB08,CCW13,SpS14,COY14,CMN15b}).

Recently there has been a growing conviction that black holes might have an intrinsically new description (known as quantum $N$-portrait), alternative and complementary to what one can deduce from a semiclassical, geometric  arguments. Theoretical evidence is supporting the idea that a black hole is a bound state or better a condensate of $N$  gravitons, whose quantum interaction strength is $1/N$ \cite{DvG13b,DvG14}.  This conjecture is reminiscent of the fuzzball proposal for black holes, whose entropy is described in terms of a bound  (BPS) state  of parallel $D-p$ branes \cite{LuM01a,LuM01b,LuM01c,Mat05,SkT08}.  The parameter $N$ is related to the pixelization of the event horizon in fundamental qubits, as it emerges from  the black hole holographic entropy \cite{Bek73,NiS14} 
\begin{equation}
N\sim\frac{r_+^2}{\ellp^2},
\label{eq:quantrule}
\end{equation}
where $r_+$ is the size of the black hole and $\ellp$ is the Planck length. Accordingly one finds $r_+\sim \sqrt{N}\ellp$ and the black holes mass $M\sim N(\frac{1}{\sqrt{N}\ellp})=\sqrt{N}\Mpl$, with $\Mpl$ the Planck mass. In this picture, the Hawking radiation corresponds to the quantum depletion of the condensate that produces a thermal spectrum with $T\sim 1/\sqrt{N}$ \cite{DvG13}. 
The black-hole-condensate is supposed to be formed in an extreme energy graviton scattering process  $2\to N$ \cite{DGI15}.  More precisely the black hole is an intermediate state between two asymptotic states, $|\mathrm{in}\rangle$ and $|\mathrm{out}\rangle$.  Both quantum field theory and string theory calculations in the large $N$ limit confirm that each final state in the reverse process $N\to 2$ is exponentially suppressed by a factor depending on the black hole entropy $S\sim N$ \cite{DGI15}.  The above  scenario is supposed to affect all  black hole formation mechanisms. These also include the gravitational collapse that in its final stages concerns matter and energy interacting at length scales of the order of the Planck length, $\ellp$. As a result, the quantum $N$-portrait efficiently describes black holes in all mass regimes - black hole macroquantumness  \cite{DvG12b} - including those one typically encounters in astrophysics.

The quantum $N$-portrait proposal has generated a considerable follow-up work, most notably in relation to the so called ``horizon wave function'' (HWF) formalism. The latter is a formulation in which one attributes to any quantum mechanical particle a wave function for the gravitational radius \cite{Cas13,CaS14,CMS14,CMS15,CMS15b,CCG15}. In doing so, one obtain probabilistic conditions to distinguish black holes from elementary particles (for a recent review see \cite{CGM15b}). As a by product, the HWF naturally encodes the generalized uncertainty principle for the particle position and the black hole decay rate.

Interestingly, the black hole quantum $N$-portrait helps to understand the very nature of gravity, that is \textit{self-complete} in the ultraviolet regime. This means that gravity prevents us from probing length scales shorter than $\ellp$. 
 In a trans-Planckian center of mass energy scattering, $\sqrt{s}=M\gg \Mpl$, all states ``classicalize'', \textit{i.e.}, they conceal to form a classical black hole \cite{Ven86,GrM88,ACV89,KPP90,Wit96,Sus02,BaF99,Ban03,DFG11,DvG12,MuN12,AuS13,AuS13b,Adl10,Car16,LaC15}. Both quantum field theory and string theory calculations in the large $N$ limit show that the black hole configuration dominates over other non-black hole classical states \cite{DGI15}.
The higher is the energy $\sqrt{s}$, the higher is the occupation number $N$ of non-propagating gravitons and the larger are the probed length scales. Accordingly also the problem of gravity bad short distance behavior, \textit{i.e.} curvature singularities, is ultimately circumvented. 

\begin{figure}[t!]
\begin{center}
\includegraphics[width=0.45\textwidth]{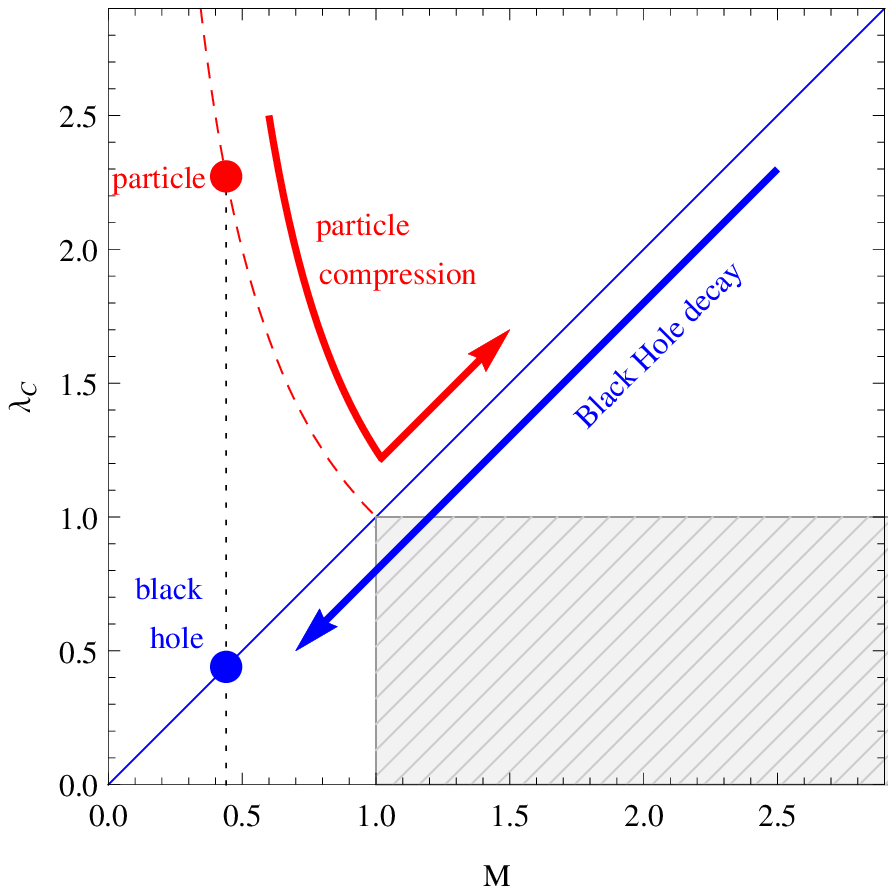} 
\includegraphics[width=0.45\textwidth]{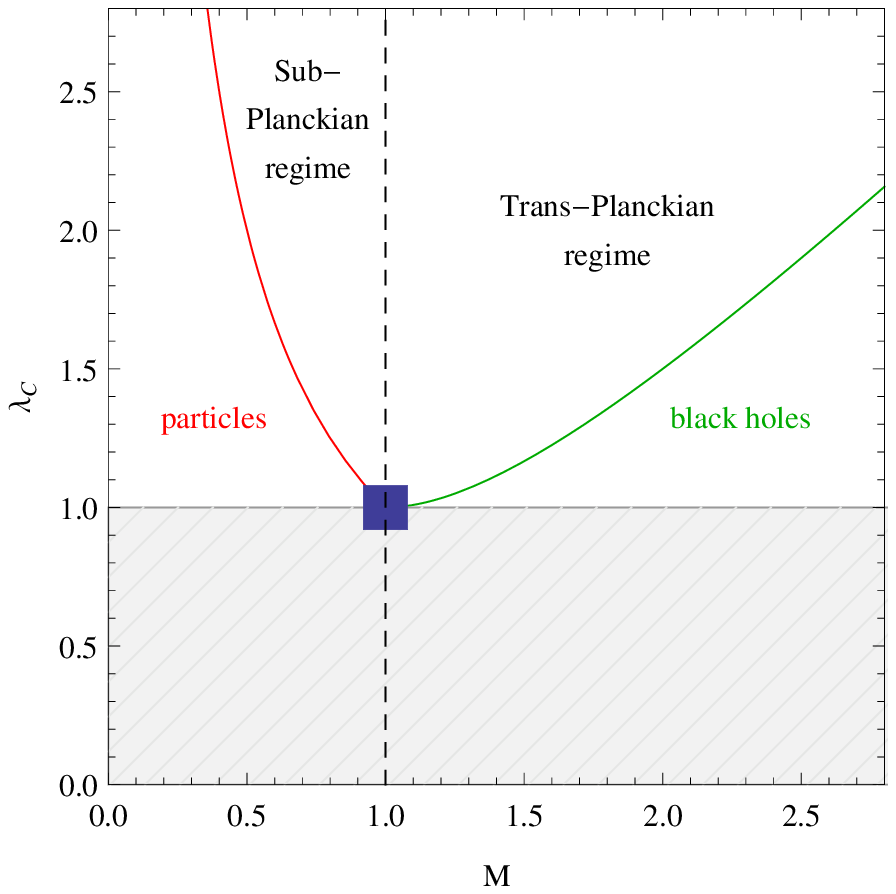}
\caption{\label{fig:bhdecay} Size vs Energy relation in Planck units. Left: The size of particles, estimated by the Compton wavelength $\lambda_{\rm C} \sim M^{-1}$, compared to the size of black holes, determined by Schwarzschild radius $r_+ \sim M$. The shaded area is inaccessible. At sub-Planckian energy scales, a length scale ambiguity arises. Right: the self-complete gravity paradigm proposes a solution.}
\end{center}
\end{figure}

Within the black hole quantum $N$-portrait paradigm, one recovers, in the large $N$-limit, the geometric picture of black holes in terms of the Schwarzschild metric \cite{DvG13,DGL15}.  On the ground of mere geometric considerations, the Schwarzschild metric is, however, in conflict with the self-complete character of gravity. Solutions of Einstein equations can exist irrespective of any lower bound for the black hole mass. For $M<\Mpl$, there is a scale ambiguity, due to the coexistence of Compton wavelengths and Schwarzschild radii. In principle, a Schwarzschild black hole can probe sub-Planckian length scales by decaying through the Hawking radiation (see Fig. \ref{fig:bhdecay} left). Of course, this scenario cannot be fully trusted due to a breakdown of the semiclassical description at the Planck scale. One expects that in the strong coupling regime, gravity actually deviates from Einstein gravity. To this purpose, an array of formulations have been proposed to consistently modify the Einstein-Hilbert action, \textit{e.g.},
\begin{equation}
 \int {\cal R}{\sqrt {-g}}\,\mathrm {d} ^{4}x\; \rightarrow 
 \int  f({\cal R},\, \Box,\, \dots)\, {\sqrt {-g}}\,\mathrm {d} ^{4}x\;,
\end{equation}
where $\Box=\nabla_\mu\nabla^\mu$ is the covariant D'Alembertian and dots stand for higher derivative terms (for a review see \cite{CaD11}).
  Accordingly, the geometric picture is also supposed to depart from the Schwarzschild geometry. In the absence of further indications from the quantum $N$-portrait beyond the large $N$ limit, one can only rely on ultraviolet improved gravity theories to draw scenarios of black holes close to the Planck scale.

Given such a background, a new metric, obtained by a process of black hole engineering, has recently been proposed in order to capture all the \textit{desiderata} of the quantum $N$-portrait for any $N$ \cite{NiS12}. In a nutshell, such a metric, known as holographic metric, enjoys the following properties:
\begin{enumerate}[i)]
\setlength{\itemsep}{-\parsep}
\item the metric admits an extremal configuration for a black hole mass $M_{\rm e}=\Mpl$ and radius $r_{\rm e}=\ellp$;
\item the extremal configuration represents the fundamental qubit of the system; any larger black holes are governed by a unique universal parameter $N$, which is set via a holographic relation similar to that between voxels and pixels \cite{NiS14};
\item the metric coincides with the Schwarzschild metric away from the Planck scale, \textit{i.e.}, in the $N\gg1$ limit.
\end{enumerate}
Notably the holographic metric fulfils the self-complete gravity paradigm. At the Planck scale one has a single graviton system represented by an extremal black hole configuration. The evaporation stops there and it is no longer possible to probe distances below the Planck length. Accordingly the ambiguity between particles and black holes for $M<\Mpl$ is removed (see Fig. \ref{fig:bhdecay} - right). At length scales $1/M<\ellp$, the geometric picture of the spacetime virtually ceases to exist, being inaccessible both via a particle compression or a black hole decay.

In this paper we  present the higher-dimensional extension of the holographic metric and we address the issue of the presence of a negative cosmological constant term. The paper is organized as follows: in Section  \ref{sec:recap}, we recap the properties of the holographic metric and we provide an alternative derivation based on nonlocal gravity equations \cite{MMN11,IMN13,Mod12a}; in Section \ref{sec:extra} we extend our findings to the case of large extra-dimensions; in Section \ref{sec:ads}, we accommodate the metric in a Anti-deSitter (AdS) background and we analyze the related thermodynamics. Finally in Section \ref{sec:concl}, we draw our conclusions.

\section{Holographic metric in $(3+1)$-dimensions}
\label{sec:recap}

In this section we recap the properties of the holographic metric. Following the lines in \cite{NiS12}, we consider the energy density for a point particle in spherical coordinates as
\begin{equation}
\label{eq:rho3ddelta}
\rho_{\rm p}(r) =
\frac{M}{4\pi r^2} \delta (r) \, .
\end{equation}
As far as $M<\Mpl$, one does not expect that the spacetime will significantly deviate from Minkowski space. Being particles characterized by a size $1/M$,\footnote[1]{As noted in \cite{DFG11}, one refers to distances and energies measured in the center of mass reference frame as seen from an ADM observer at spatial infinity.}
 the curvature turns to be at the most ${\cal R}\sim M^2(M/\Mpl)^2\ll \Mpl^2$.
On the other hand, Einstein's equations allow for gravitational collapse for any $M$.  As a result, one has that \eqref{eq:rho3ddelta} can also accommodate black holes for any sub-Planckian mass $M$.   %

Before proceeding we recall that \eqref{eq:rho3ddelta} can be written in terms of the Heaviside function, being $\delta(r)=\d\Theta(r)/\d r$. 
The sharp step function is, however, believed to be modified by wild quantum disturbances at the Planck scale. This suggests that $\Theta(r)$ has to be replaced by a smooth function $h(r)$, accounting for the average of the quantum gravity fluctuations. The exact profile of the $h(r)$ is, in general, not known. In the absence of theoretical indications or experimental constraints, we can reduce the class of profiles of $h(r)$ only by invoking some guiding principles, \textit{e.g.}, the self-complete character of gravity. In doing so, we aim to remove the particle-black hole ambiguity of the Schwarzschild geometry in the sub-Planckian regime.  To reach this goal, we need to derive a metric admitting an extremal configuration, whose size (radius and mass) sets the characteristic scales of gravity. 

We proceed by keeping $h(r)$ general and we express \eqref{eq:rho3ddelta} as
\begin{equation}
\label{eq:rho3d}
\rho(r) =
\frac{M}{4\pi r^2} \dd{h(r)}{r}.
\end{equation}
The above energy density is no longer infinitesimally narrow. This means that there exists a non-vanishing energy momentum tensor, whose time-time mixed component, $\mathfrak{T}^{\ 0}_0$, is determined by \eqref{eq:rho3d}. All non-classical phenomena are incorporated in such an effective energy momentum tensor. The procedure allows for the derivation of a Schwarzschild-like geometry with a mass profile $m(r)$ which collects all deviations from Einstein gravity, \textit{i.e.}, 
\begin{equation}
\d s^2 = 
- \left( 1 - \frac{2 G m(r)}{r} \right) \d t^2
+ \left( 1 - \frac{2 G m(r)}{r} \right)^{-1} \d r^2
+ r^2 \d \Omega^2.
\label{eq:line4d}
\end{equation} 
Here the function $m(r)$ is given by
\begin{equation}\label{eq:mass-3d}
m(r) = 4\pi \int \d r ~ r^2 ~ \rho(r)
= M \int \d r~ \dd{h(r)}r \rightarrow m(r)= M~h(r)
\end{equation}
where the boundary condition $m(r)\to M$ as $r\to \infty$ has been used to fix the integration constant.
The conservation equation $\nabla_\mu \mathfrak{T}^{\mu\nu}=0$ and the ``clean black hole condition'' $g_{00}=-g_{rr}^{-1}$ univocally determine the effective energy momentum tensor $\mathfrak{T}_\mu^{\ \nu} = \diag\left(-\rho,\ p_r,\ p_\perp,\ p_\perp\right)$ with equation of state $p_r=-\rho$ and  angular pressure $p_\perp = p_r + r \partial_r p_r / 2$.  The profile of $\mathfrak{T}^{\mu\nu}$ corresponds to the case of an anisotropic fluid peaked at the origin and leads to a local violation of energy conditions. Energy momentum tensors of this kind have already been employed in a variety of black hole models aiming to improve classical spacetime geometries, such as vacuum nonsingular black holes \cite{Dym92,Mag99,Gia02,MbK05,Hay06,BMM13,Nev15}, noncommutative geometry inspired black holes \cite{NSS06b,Nic09,NiS10}, nonlocal gravity black holes \cite{MMN11,Nic12}, generalized uncertainty principle inspired black holes \cite{IMN13}, nonlinear electric source black holes \cite{ABG98,ABG99,ABG99b,ABG00,Cul15}.

\begin{figure}[t!]
\includegraphics[width=0.5\textwidth]{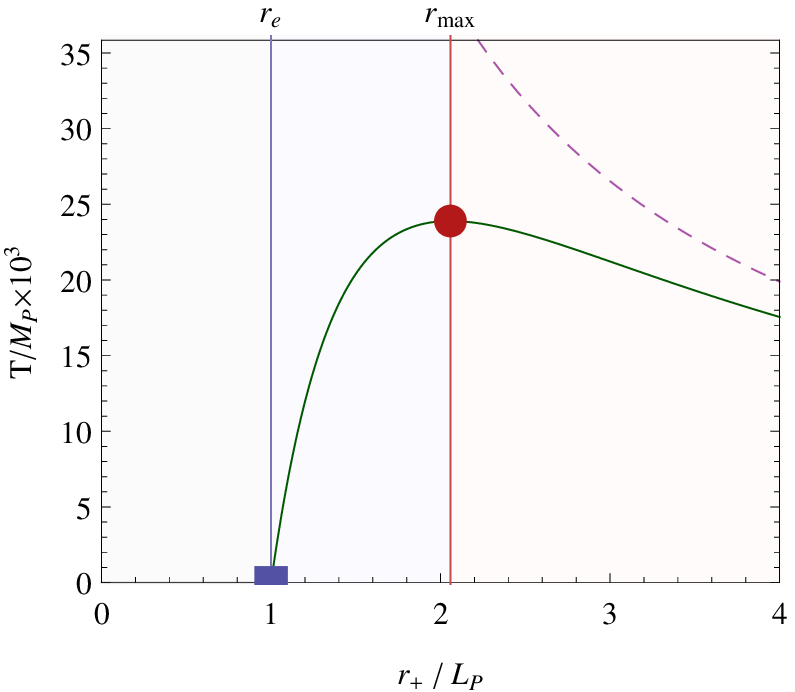}%
\includegraphics[width=0.5\textwidth]{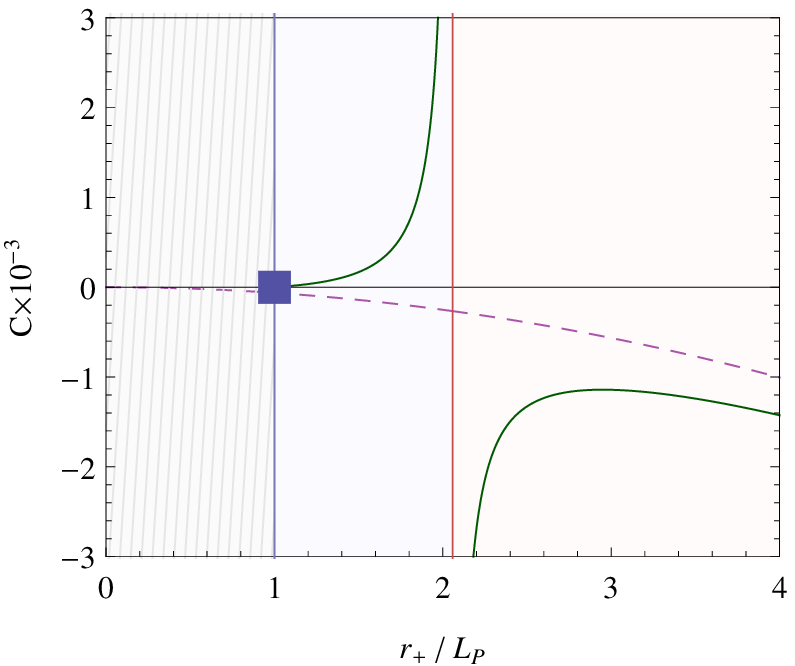}%
\label{fig:TCS3d}%
\caption{Left: Temperature of the holographic metric in $(3+1)$-dimensions; Right: heat capacity $C\equiv {\rm d}M/{\rm d}T$. The shades area is inaccessible $r < r_{\rm e}=\ellp$. The asymptote in the heat capacity occurs at the maximum temperature. The positive heat capacity phase corresponds to the cool down phase also known as ``SCRAM phase''. }
\end{figure}

From \eqref{eq:mass-3d}, one can see that all metric modifications are now expressed in terms of the function $h(r)$. For consistency there should be a characteristic length scale such that for larger distances the metric in \eqref{eq:line4d} reproduces the standard Schwarzschild solution. Since in gravity there exists just a unique length scale, we conclude that $h(r)\to 1$ for $r\gg \ellp$. The profile of $h(r)$ has also to obey the tenets of the self-complete gravity paradigm. To reach this goal, one requires that $h(r)$ allows for horizon extremisation when $M=\Mpl$. The net result is a metric that matches the Schwarzschild metric at large distances and admits a Planckian extremal black hole configuration. The most compact profile fulfilling the above requirement is  
\begin{equation}\label{eq:h-3d}
h(r) = 1 - \frac{\ellp^2}{r^2 + \ellp^2}.
\end{equation}
Accordingly the holographic metric reads \cite{NiS12}
\begin{equation}
\d s^2 = 
- \left( 1 -  \frac{2 \ellp^2 \ M \ r\ }{r^2+\ellp^2} \right) \d t^2
+ \left( 1 -  \frac{2 \ellp^2 \ M \ r\ }{r^2+\ellp^2} \right)^{-1} \d r^2
+ r^2 \d \Omega^2.
\label{eq:line4dholo}
\end{equation} 
This metric enjoys the property of having a couple of horizons
\begin{equation}
r_\pm=\ellp^2\left(M\pm\sqrt{M^2-\Mpl^2}\right),
\end{equation} 
 with $r_-<\ellp$ and $r_+>\ellp$, that coalesce in the extremal configuration, $r_{\rm e}\equiv r_+=r_-=\ellp$, for $M=\Mpl$.
Extremal black holes are zero temperature states, a fact that is evident by analyzing the thermodynamic of \eqref{eq:line4dholo}. By evaluating the black hole surface gravity $\kappa$, one obtains
\begin{align}
\label{eq:T3d}
T &=
\frac{1}{4\pi r_+} 
\left( 1 - \frac{2 \ellp^2}{r_+^2 + \ellp^2} \right).
\end{align}
The profile of the temperature (see Fig. \ref{fig:TCS3d}) reveals that the hole undergoes a phase transition from a negative heat capacity warming to a positive heat capacity cooling, also known as black hole SCRAM \cite{Nic09}. At $r_+=r_{\rm e}=\ellp$, $M=\Mpl$, the black hole decay virtually halts and  for $M<\Mpl$ no event horizons form. Interestingly the curvature singularity cannot be probed either by a particle compression or black hole decay, in agreement to what sought within the self-complete gravity paradigm (see Fig. \ref{fig:bhdecay}).  

The calculation of the black hole entropy
\begin{align}
\label{eq:S3d}
S &= \int_{\ellp}^{r_+} \frac{\d M}{T}
= \frac{\pi}{\ellp^2} (r_+^2 - \ellp^2)
+ 2\pi \ln \left( \frac{r_+}{\ellp} \right)
\end{align}
discloses further interesting features of the holographic metric. The non-classical corrections have logarithm dependence as largely expected in all the major approaches to quantum gravity. Furthermore one can use the quantization rule \eqref{eq:quantrule} to express the entropy in terms of the parameter $N$ as
\begin{align}
S &=  \pi \left( N -1\right)
+ \pi \ln \left( N \right).
\end{align}
In other words the extremal configuration, $N=1$, represents the basic information capacity of the system, the holographic representation of the black hole quantum constituent, \textit{i.e.}, the graviton. In the large $N$ regime, the black hole entropy is just a multiple of the fundamental area $r_{\rm e}^2$.

Interestingly the parameter $N$ governs all metric corrections. For $r>2GM$, we can re-write the line element \eqref{eq:line4d} as
\begin{equation}
\d s^2 \approx 
- \left[ 1 -  \frac{2 G M }{r}\left(1+\frac{1}{N}\right) \right] \d t^2
+ \left[ 1 -  \frac{2 GM }{r}\left(1+\frac{1}{N}\right) \right]^{-1} \d r^2
+ r^2 \d \Omega^2.
\label{eq:line4dN}
\end{equation} 
where non-classical corrections die off as $1/N$ in agreement with the quantum $N$-portrait \cite{DvG13}.  We note that the above $1/N$ term also matches the quantum mechanical corrections to the horizon radius obtained by means of the generalized uncertainty principle \cite{CMN15}. Quantum mechanical effects at the horizon are a key feature of the fuzzball proposal too \cite{Mat05,SkT08}, a fact that is signalling the convergence towards a model independent scenario.

\subsection{Derivation in terms of non-local gravity actions}
In the recent years there has been a lot of interest in modifications of Einstein gravity both in the ultraviolet and infrared regimes \cite{CaD11}. A notable example is offered by the class of non-local gravity deformations, theories that exhibit an infinite number of derivative terms \cite{MMN11,Nic12,IMN13,Mod12a,Mof11,Kra87,Tom97,Bar03,Bar05,Bar12,BGK11,
CMN14,GHS10}.

In order to derive the holographic metric, we consider the following non-local action \cite{Bar03,Bar05,Bar12} in place of the conventional Einstein-Hilbert action
\begin{eqnarray}
I = - \frac{1}{16\pi G} \int {\rm d}^4 x \sqrt{ - g} \, 
 G^{\mu \nu}
 \, \frac{\mathcal{A} ( \Box/\mu^2)}{\Box} R_{\mu \nu}.
\label{eq:nlaction}
\end{eqnarray}
where $\mu$ is the non-local gravity scale and $\mathcal{A}$ is the non-local operator.
The above action is a truncation of the action proposed in \cite{Mod12a,Kra87,Tom97}. To a first approximation non-local actions  lead to the same field equations \cite{MMN11,Mof11,GHS10}
\begin{eqnarray}
\mathcal{A} \left( \Box/\mu^2\right) \left( R_{\mu \nu} -\frac{1}{2} g_{\mu \nu} R \right) 
+ O(R_{\mu \nu}^2) 
= 
8 \pi G T_{\mu \nu}. 
\label{eq:nleqns1}
\end{eqnarray}
Here $T_{\mu \nu}$ is derived from an action describing ordinary matter, where the gravity part is affected by non-local effects.  Following the line of reasoning in \cite{MMN11,IMN13,Mof11,GHS10}, we neglect higher order corrections and we cast the non-local equations 
 in the following form
\begin{eqnarray}
 R_{\mu \nu} -\frac{1}{2} g_{\mu \nu} R
= 8 \pi G \ \mathfrak{T}_{\mu \nu},
\end{eqnarray}
where the non-local energy momentum tensor, $\mathfrak{T}_{\mu \nu}\equiv \mathcal{A}^{-1} \left( \Box /\mu^2\right) T_{\mu \nu}$, is coupled to the standard Einstein tensor.

We can exploit the above equations to reproduce the effective energy momentum tensor we employed for the holographic metric. We start the calculation by considering a static source. We recall that the standard Schwarzschild geometry can be derived by considering a static point-like particle sitting at the origin with energy profile given by \eqref{eq:rho3ddelta}. As a result, any non-local energy density deformation is of the form
\begin{eqnarray}
\mathfrak{T}_0^{\ 0} &=& M \C A^{-1}\left(\square/\mu^2\right) \delta^3(\vec x)
\\
&=& M \C A^{-1}\left(\Delta/\mu^2\right)\delta^3(\vec x)\nonumber
\end{eqnarray}
where $\Box$ becomes the Laplace operator $\Delta$ in case of static sources.
Apart from being entire functions in momentum space $\C A (-p^2)$, there are no additional, theoretical or experimental constraints on the profile of the operators. One can  invoke some principles to postulate a specific profile. 
As a consequence we can select a profile of $\C A$ to reproduce the self-complete character of gravity encoded in the holographic metric.
This is equivalent to solving  the following equation in term of 
$\C A$
\begin{eqnarray}
\label{eq:condA}
\left(\frac{1}{2\pi |\vec x|}\right)\frac{\ellp^2}{\left(|\vec x|^2+\ellp^2\right)^2} =
 \C A^{-1}\left(\Delta/\mu^2\right)\delta^3(\vec x)
\end{eqnarray}
with $\mu\sim \Mpl$.

Following the procedure in \cite{MMN11,Mof11,GHS10}, we can Fourier transform the equation \eqref{eq:condA} and algebraically determine the profile of ${\cal A}(p)$ where $p=|\vec{p}|$. The result reads
\begin{equation}
\label{eq:Asolution}
\C A^{-1}(p) =
 \frac{1}{2} \left[
e^{p/\mu} {\rm E}_1(p/\mu) \left( \frac{1}{p/\mu} - 1 \right)
+
e^{-p/\mu} {\rm Ei}(p/\mu) \left( \frac{1}{p/\mu} + 1 \right)
\right]
\end{equation}
where
\begin{equation}
{\rm Ei}(x) = \int_{-\infty}^x \frac{e^t}{t} \d t
\quad \text{and} \quad
{\rm E}_1(x) = \int_x^\infty \frac{e^{-t}}{t} \d t .
\end{equation}
The above function can be expanded in power series and written back into coordinate space by means of the following Schwinger representation for operators
\begin{equation}
\C O^{\alpha}=\frac{1}{\Gamma(-\alpha)}\int_0^\infty\frac{\d s}{s}\ s^{-\alpha} \ e^{-s\ \C O}
\end{equation}

In particular $p$ turns out to be dual to $(-\Delta)^{1/2}$.
We note that the function $\C A^{-1}$ in \eqref{eq:Asolution} is an entire function, it admits the inverse function $\C A$, and can be Taylor expanded for $p\ll \mu$.
As a result one finds that the operator in \eqref{eq:condA} can be written as a power series: 
\begin{align}
\C A^{-1} (
\Box) \approx &\   1 + \left[\frac{\gamma}{3}-\frac{4}{9} + \frac{1}{6}\ln\left(\frac{\Box}{\mu^2}\right)\right]\left(\frac{\Box}{\mu^2}\right)+\frac{1}{60} \left[3\gamma-4+\ln\left(\frac{\Box}{\mu^2}\right)\right]\left(\frac{\Box}{\mu^2}\right)^2 \nonumber\\
 &
 + O(\Box^3/
\mu^6)
\end{align}
where the logarithm of an operator is defined by  $\exp(\ln\C O)=1$. We conclude that the Einstein gravity is consistently recovered in the limit 
\begin{equation}
 \lim_{\mu\to\infty}\C A\left(\Box/\mu^2\right)= 1.
 \end{equation}

The procedure can be extended by including higher order corrections in the equations and further terms in the action. Such terms are important only at the Planck scale or beyond, \textit{i.e.},  at energies where the very concept of line element and the geometric description of the black hole actually breaks down. In principle one could exploit forthcoming indications from the quantum $N$-portrait (possibly away from the large $N$ limit) to get information about such additional terms as well as stricter constraints on $\C A$. As far as we are concerned of energies below the Planck scale, the first order approximation efficiently works. We also recall that the extremal configuration is an asyptotic state. This means that the presented profile can be safely used for all practical purposes.

\section{Higher dimensional holographic metric}
\label{sec:extra}

In the past two decades there has been a considerable attention about the so called terascale quantum gravity, a phenomenological repercussion of large extra dimensions \cite{AAD98,ADD98,ADD99} and  brane-world scenario \cite{RaS99a,RaS99b}. The later are two paradigms proposed to tackle the hierarchy problem, \textit{i.e.}, the huge discrepancy between the Planck scale, $\Mpl\sim 10^{19}$ GeV and the electroweak scale $\Lambda_{\rm ew}\sim 10^{2}$ Gev. The common feature of such scenarios is the possibility of having a new fundamental scale  around the terascale, $M_\ast\sim 1$ TeV, by allowing the spacetime to have additional spatial dimensions. A lower fundamental scale is equivalent to having a stronger gravitational interaction and a rich quantum gravity phenomenology in particle collisions or in high energy cosmic ray showers. The latter includes the colliding particle gravitational collapse in microscopic black holes \cite{BaF99,ADM98,DiL01,GiT02,FeS02,AFH02,KRT02} (for reviews on the topics see \textit{e.g.} \cite{Lan02,Cav03,Kan04,Hos04,CaS06,Ble07,
Win07,BlN10,Cal10a,Par12,KaW15}). 
Although experimental investigations up to $\sqrt{s}=8$ TeV put severe constraints on extradimensional models and black hole production \cite{CMS15c,ATL15}, extradimensions remain the only viable tool to address the incompleteness of the Standard Model (see \cite{MNS12,BlN14} for some comments about the non-observation of terascale quantum gravity).  In addition extradimensions have been proposed in formulations, that being alternative to the large-extradimensions and the brane world scenario,  might not suffer from current experimental constraints \cite{ACD01,Gog99,Gog00,Gog02}. This is equivalent to saying, that apart from the mathematical importance, the study of higher dimensional black holes might still have phenomenological repercussions in high energy physics. 

On the ground of results in the previous section we present the higher dimensional extension of the holographic metric. For brevity we display the property of the metric in just one set up, the large extradimension  model. We introduce the fundamental mass $M_\ast$ by integrating out the volume of the extra dimensions $V_n$,
\begin{equation}
\Mpl^2 = C_n V_n M_*^{n+2},
\end{equation}
where $n$ is the number of extra dimensions, $D=d+1=4+n$ is total number of dimensions 
and $C_n$ is a dimensionless prefactor of the order of unity.
In case of toroidal compactification the volume is 
$V_n=(2\pi R_{\rm c})^n$ where $R_{\rm c}$ is the compactification radius of the extra dimensions.

The system has a length scale defined by $L_*=1/M_*$, that will be the order of magnitude of the size of any black hole of mass $M\sim M_\ast$.
To neglect boundary effects  we assume the fundamental scale be much smaller than the compactification radius $R_{\rm c} \gg L_*$. Accordingly an hyperspherical black hole solution consistently describes the spacetime geometry irrespective of brane effects.

We proceed by considering a modified Schwarzschild-Tangherlini energy density as for the four dimensional case, \textit{i.e.},
\begin{equation}
m(r) = \int \d^{d} x ~\rho(
\vec{x})
=  \Omega_{d-1} \int \d r ~ r^{d-1} \rho(r)
\stackrel{!}{=} M h_n(r).
\end{equation}
where $\Omega_{d-1}$ is the surface of the unit $(d-1)$-sphere, given by $\Omega_{d-1}=2\pi^{d/2}/\Gamma(d/2)$.
From the equation above we can write
\begin{equation}\label{eq:densityN}
\rho(r) = \frac{M}{\Omega_{2+n} r^{2+n}} \dd{h_n(r)}{r}.
\end{equation}
where the function $h_n(r)$ is higher dimensional version of the function $h(r)$ in Section \ref{sec:recap}. To solve the $D$-dimensional Einstein's  equations, we assume a static, neutral metric of the kind
\begin{equation}
\d s^2 = 
- (1 +2 \Phi(r)) \d t^2 
+ (1 +2\Phi(r))^{-1} \d t^2 
+ r^2 \d \Omega_{2+n}^2.
\end{equation}
with 
\begin{equation}
\d\Omega_{n+2}^{2}
=\d\vartheta_{n+1}^2+\sin^2\vartheta_{n+1}\left(\d\vartheta_n^2+\sin^2\vartheta_n^2\left(\dots +
\sin^2\vartheta_2(\d\vartheta_1^2+\sin^2\vartheta_1\d\varphi^2)\dots\right)\right)
\end{equation}
The integration of the equations leads to the Newton's potential expressed in terms of $h_n(r)$:
\begin{equation}
\Phi(r) =-
\frac {8\pi}{(2+n)\Omega_{n+2}}
\frac M{M_*^{n+2}} \frac{h_n(r)}{r^{1+n}}.
\end{equation}

\begin{figure}[t!]
\begin{center}
\includegraphics[width=0.46\textwidth]{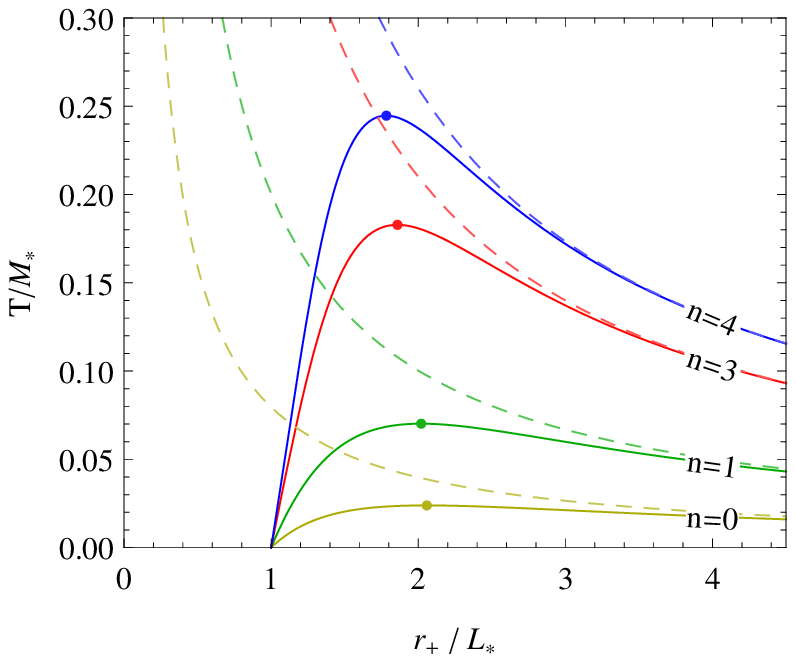} 
\includegraphics[width=0.46\textwidth]{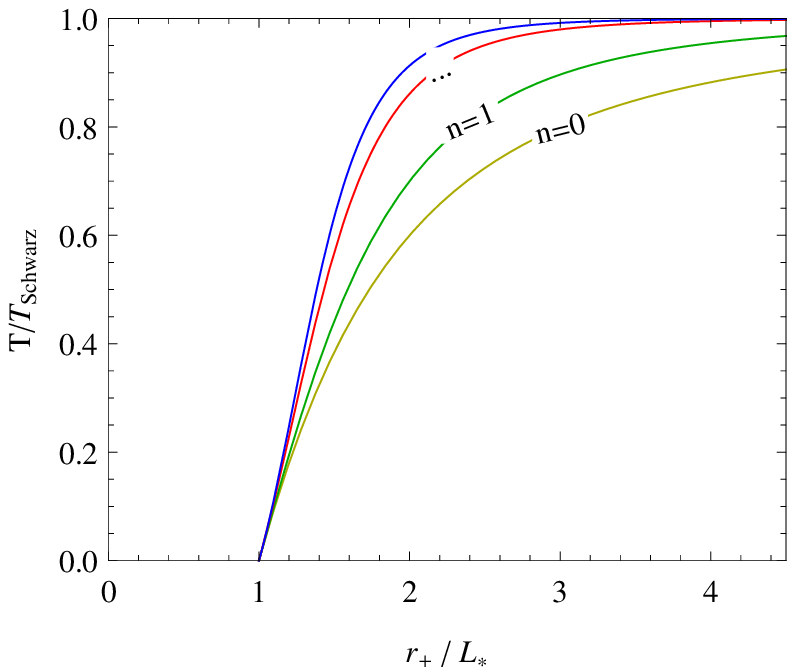}
\caption{\label{fig:thermo-n}Left: Temperatures of the higher dimensional holographic metric ($n$ is the number of spatial dimensions); Right: ration between the holographic metric temperature and that of the Schwarzschild-Tangherlini temperature black hole, $T_{\rm Schwarz}= (1+n)/(4\pi r_+)$. 
}
\end{center}
\end{figure}

From the above line element one can calculate relevant thermodynamic quantities. For instance the black hole temperature reads
\begin{align}
T &= \frac{n+1}{4\pi r_+} \left( 1-\frac{r_+}{n+1} \frac{h_n'(r_+)}{h_n(r_+)} \right)
\end{align}
while the generic profile of the entropy is
\begin{align}
S &=
\frac{(n+2)}{4}\ \Omega_{n+2} M_*^{n+2}
\int \d r_+ \frac{r_+^{1+n}}{h_n(r_+)}
\end{align}

\begin{table}[t!]
\begin{center}
\begin{tabular}{lcccccccc}
\firsthline
 $n$ & 0 & 1 & 2 & 3 & 4 & 5 & 6 & 7 \\
   \hline
 $r_{\rm max}/L_*$ & 2.06 & 1.60 & 1.48 & 1.41 & 1.36 & 1.33 & 1.30
   & 1.28 \\
 $T_{\rm max}/M_*$ & 0.024 & 0.07 & 0.12 & 0.18 & 0.25 & 0.31 & 0.38 & 0.44 \\
   \hline
\end{tabular}
\end{center}
\caption{\label{table:holo-n} Maximum radii $r_{\rm max}$ and maximum temperatures $T_{\rm max}\equiv T(r_{\rm max})$ of the holographic black hole in $(3+n)$  spatial dimensions.} 
\end{table}

\begin{figure}[t!]
\begin{center}
\includegraphics[width=0.75\textwidth]{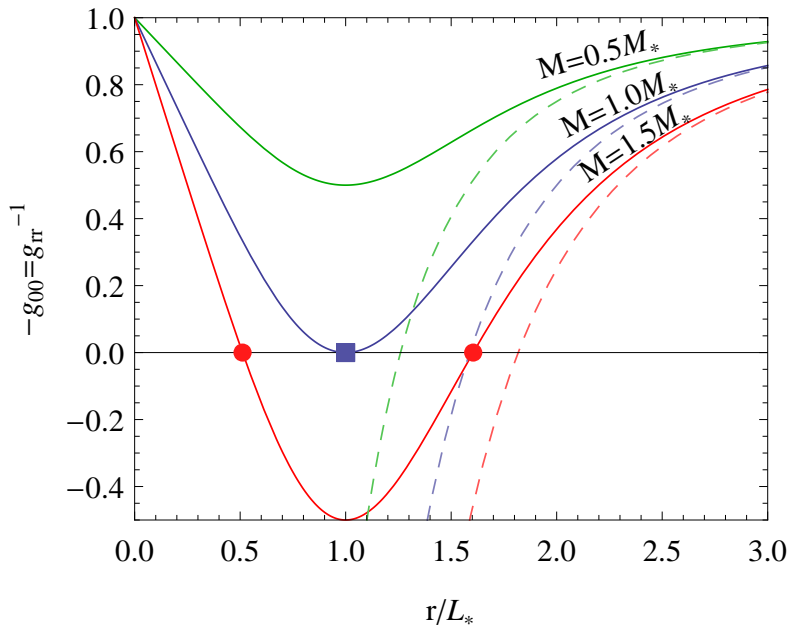} 
\caption{\label{fig:g00-n}The function $-g_{00}$ vs $r$ for $n=2$. The three curves corresponds to three different values of the mass parameter $M$. 
}
\end{center}
\end{figure}

The actual profile of the holographic metric in higher dimensions can be found by means of entropic arguments.
By requiring that entropy corrections have a logarithmic dependence in any dimensions, \textit{i.e.}, $S \stackrel{!}{\sim} \int \d r_+ / r_+$, we can postulate that the function $h_n(r)$ in higher dimensions has to be:
\begin{equation}\label{eq:mass-nd}
h_n(r) =  \frac{r^{2+n}}{r^{2+n} + L_*^{2+n}}.
\end{equation}
To check that this is the correct profile we need to verify whether the conditions for a remnant at the fundamental scale are fulfilled. To do this, we calculate the radius of extremal configuration, $r_{\rm e}$, \textit{i.e.}, the solution of the following system of equations 
\begin{equation}
\begin{cases} g_{00}(r)=0 \\ \frac{\d}{\d r}\ g_{00}(r)=0\end{cases} 
\end{equation}
We notice that the function $|g_{00}(r)|\to 1$  for both $r\gg L_\ast$ and $r\sim L_\ast$. This means that it must admit at least a stationary point for intermediate values of $r$. From Fig.~\ref{fig:g00-n} one can see that $g_{00}$ describes a family of curves labeled by the mass parameter $M$.  For $M>M_\ast$, the curve $g_{00}$ intersects the $r$-axis two times, \textit{i.e.}, in $r_\pm$. For $M<M_\ast$, the curve $g_{00}$ never intersects the $r$-axis and no horizons form. Finally for $M=M_\ast$, the curve $g_{00}$ has a double zero in $r_{\rm e}=L_\ast$. 

The horizon extremisation at $L_\ast$ guarantees the self-complete character of the metric, as it is evident by analyzing the profile of the temperature in Fig~\ref{fig:TCS3d}. As in the four dimensional case, the black hole reaches a maximum temperature $T_{\rm max}$ before  ``scramming'' down towards a zero temperature thermodynamic limit (see Fig. \ref{fig:thermo-n}). This implies that the holographic horizon is colder with respect to the Schwarzchild black hole of the same size and does not suffer of relevant back reaction. Up to $n=7$, the temperature/mass ratio is $T/M< T_{\rm max}/M_\ast\leq 0.44$ during all the evaporation process  (see Tab. \ref{table:holo-n}).  This lets us conclude that, in analogy to what found in \cite{NiW11}, the holographic black hole tends to emit softer particles mainly on the brane, \textit{i.e.}, with a reduced bulk emission with respect to Schwarzschild black holes with the same mass. It would be interesting to study the rotating version of the holographic metric and understand how the reduction of its spectrum competes with superradiance effects recently found to take place in the extradimensional scenario (see \cite{FrS02,FrS03})

We notice, however, that by increasing the number of extradimensions $n$, the holographic horizon becomes relatively hotter. 
The ratio $T_{\rm max}/M_\ast$ increases with $n$ (see Tab. \ref{table:holo-n}),  rather than decreasing as found in \cite{NiW11}.  This is due to the fact that the temperature of the holographic metric drops to zero at $r_e=L_\ast$ irrespective of  $n$. On the contrary models studied in \cite{NiW11} have smaller remnant radii as $n$ increases and flatter temperature profiles, with a maximum that decreases with $n$. This is not the case for the holographic metric. We conclude that this might be a peculiar feature of  self-completeness.

\section{AdS background and Hawking-Page phase transition}
\label{sec:ads}

The holographic metric in Section \ref{sec:recap} showed intriguing thermodynamic properties such logarithmic corrections to the area law.  The pixelization of the event horizon let us describe the system in terms of a unique parameter $N$, in the same fashion of what happens within the quantum $N$-portrait. In the present section, rather than black holes in asymptotically flat space, we consider the case of a non-vanishing cosmological constant. In particular we aim to study the case of a black hole in AdS space, characterized by a negative cosmological constant $\Lambda=-3/b^2$, where $b$ is the AdS radius. The AdS curvature term can equivalently be  described in terms of a fluid-like stress tensor, characterized by negative pressures and tendency to collapse. An AdS background is also a key element of the gauge-gravity duality, a correspondence between some strongly coupled field theories and (weakly curved) gravitational systems \cite{Mal99}.

In the conventional Schwarzschild-AdS geometry, the black hole has a minimum Hawking temperature $T_{\textrm{min}}$ for $r_+=r_0=b/\sqrt{3} $. 
This means that black holes have positive specific heat for $r_+>r_0$  and can be in stable equilibrium with the thermal radiation at fixed temperature.
On the other hand, at temperature below $T_{\textrm{min}}$, there is no possibility for the thermal gas to collapse and form an AdS black hole. The thermal gas is thus stable. This suggests that a thermal phase transition should occur at some temperature $T_{\textrm{HP}} \geq T_{\textrm{min}}$, with the thermal gas dominating at temperatures below $T_{\textrm{HP}}$ and the large black hole dominating above $T_{\textrm{HP}}$. This is the so-called Hawking-Page transition \cite{HaP83}.

It is therefore natural to ask what would be the impact of a metric encoding the self-complete character of gravity on Hawking-Page transition and the  thermodynamics of AdS black holes.
A non-vanishing cosmological term, however, requires some revision. The presence of thermal radiation modifies the conditions for Planckian remnants and attaining a  ultraviolet self-complete metric.  One needs to consider that the class of functions $h(r)$ in Section \ref{sec:recap} is just the limit of a larger class of functions, \textit{i.e.}, $h_b(r)\to h(r)$ for $b\to \infty$. Therefore we can postulate that the generic line element reads
\begin{eqnarray} \label{eq:metro}
\d s^2 &=& 
- \left( 1 -  \frac{2 \ellp^2 \ M }{r}\ h_b(r)+\frac{r^2}{b^2} \right) \d t^2
+ \left( 1 -  \frac{2 \ellp^2 \ M }{r}\ h_b(r)+\frac{r^2}{b^2} \right)^{-1} \d r^2 \nonumber\\
&&+ r^2 \d \Omega^2.
\end{eqnarray}
with $h_b(r)\to 1$ as $r\to\infty$.
The internal energy of the system can be obtained by the equation $g_{00}=0$ and reads
\begin{equation}
\label{eq:Mb}
M(r_+)=\frac{r_+}{2\ellp^2 \ h_b(r_+)}\left(1+\frac{r_+^2}{b^2}\right) \,.
\end{equation}
The horizon temperature can be calculated by evaluating the surface gravity 
 \begin{equation}
T=\frac{1}{4\pi r_+}\left[\left(1+\frac{r_+^2}{b^2}\right)\left(1-r_+\frac{h_b^\prime(r_+)}{h_b(r_+)}\right)+\frac{2r_+^2}{b^2}\right]
\label{eq:adstemp1}
\end{equation}
and one can check that for $h_b\approx 1$, $h_b^\prime\approx 0$, the above temperature matches the standard Schwarzschild-AdS temperature. 
To obtain the \textit{extremal} configuration we need to couple \eqref{eq:Mb} with the equation $g_{00}^\prime=0$, \textit{i.e.},
\begin{equation}
 g_{00}^\prime(r)\bigg|_{r=r_{+}}=
\left[ \frac{2M\ellp^2}{r^2}h_b(r)-\frac{2M\ellp^2}{r}h_b^\prime(r)+\frac{2r}{b^2}\right]_{r=r_{+}}=0,
\label{eq:g00bprime}
\end{equation}
and assuming that $r_+=\ellp$, one can use \eqref{eq:Mb} in \eqref{eq:g00bprime} to get
\begin{equation}
g_{00}^\prime(\ellp)=\left(1+\frac{\ellp^2}{b^2}\right)\left(3\Mpl-\frac{h_b^\prime(r_+)}{h_b(r_+)}\right)-2\Mpl=0.
\label{eq:g00bprimeplanck}
\end{equation}
We notice that for any finite $b$,  the extremal mass at the Planck scale reads
\begin{equation}
\label{eq:Mbe}
M_{\rm e}=M(\ellp)=\frac{\Mpl}{2 \ h_b(r_+)}+\frac{1}{2 \ h_b(r_+)}\frac{\ellp}{b^2}>\Mpl
\end{equation}
and equals the Planck mass only in the limit $b\to\infty$. The physical reason is due to the fact that our black hole is immersed in a thermal bath. We recall that the cosmological term emerges by coupling Einstein tensor to a energy momentum tensor $T_{\mu\nu}^{\Lambda}=-\frac{\Lambda}{8\pi G}g_{\mu\nu}$. The amount of energy  related to the cosmological term in a volume $V$ can be estimated by $E_\Lambda\sim T_0^{\ 0}V$. For $V\sim \ellp^3$ one gets $E_\Lambda\sim \ellp/b^2$, \textit{i.e.}, the additional term on the r.h.s. of \eqref{eq:Mbe}.
As a result we can only require that the remnant radius $r_{\rm e}=\ellp$ while $M_{\rm e}\to \Mpl$ only in the limit $b\to\infty$.
Using $h_b\to h$ for $b\to\infty$, we find that 
\begin{equation}
h_b(r)=\frac{r^2}{r^2+\ellp^2+\frac{4\ellp^4}{b^2-\ellp^2}}
\label{eq:hbads}
\end{equation}
is a solution of \eqref{eq:g00bprime}.
The above profile is a legitimate distribution to replace the Heaviside function. We have the limits $h_b\to 0$ as $r\to 0$ and $h_b\to 1$ as $r\to\infty$. This guarantees that the spacetime is asymptotically AdS-space.
Interestingly, for $b\to \ellp$, one has that $h_b\to 0$. This means that, contrary to the Schwarzschild-AdS case, there is no black hole and the spacetime becomes a purely thermal AdS-state. In other words there exists a minimum mass for black hole formation. If the AdS-radius is too small, the black hole cannot fit in.
Inserting the profile {\eqref{eq:hbads} in {\eqref{eq:metro}}, one has that the metric coefficient is} 
\begin{equation}
g_{00}=1 -\frac{2 M \ellp^2  r}{\ellp^2+r^2+\frac{4\ellp^4}{b^2-\ellp^2}} + \frac{r^2}{b^2} \,.
\label{eq:g00holoads}
\end{equation}
From the horizon equation $g_{00}=0$, one finds that there exists a value  $M_{\rm e}$, depending on $b$ and $\ellp$, such that: 
for $M>M_{\rm e}$ the metric admits two horizons $r_\pm$, with $r_-<\ellp$ and $r_+>\ellp$;
for $M<M_{\rm e}$ the metric has no horizons;
for $M=M_{\rm e}$ the metric admits one degenerate horizon $r_{\rm e}=r_+=r_-=\ellp$.

The extremal horizon guarantees the ultraviolet self-completeness character of the metric.The inner horizon, being smaller than the Planck length, is actually inaccessible. This facts downplays the problem of the Cauchy instability and the related mass inflation \cite{PoI89,PoI90,BaN10,BrM11} much in the same way as the problem of the curvature singularity. The self-complete character allows to hide all kinds of pathologies of the metric behind the Planck barrier.

\begin{figure}[t!]
\begin{center}
\includegraphics[width=0.75\textwidth]{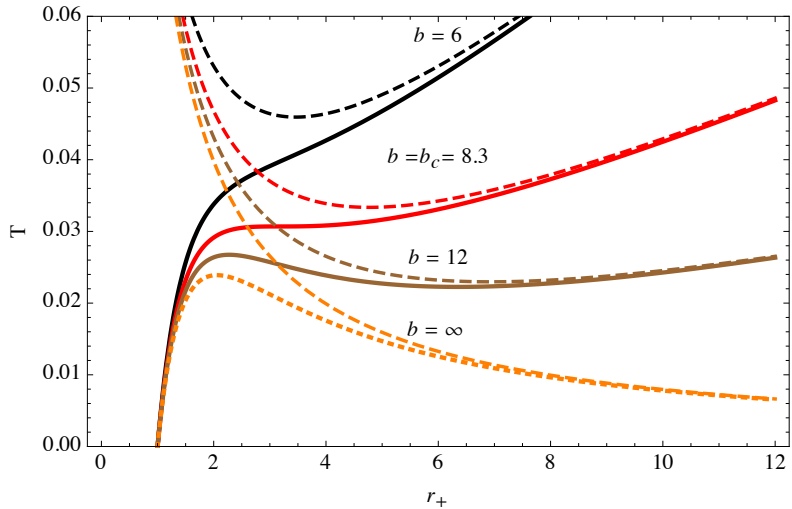}
\caption{Plot of the Hawking temperature \eqref{eq:holoadstemp} as function of the horizon radius $r_+$ with $\ellp=1$. The temperature goes to zero for $r_+=\ellp$. The dashed line shows the behaviour of the standard Schwarzschild-AdS temperature. The $b\to\infty$ case corresponds to the temperature of the asymptotically flat holographic metric. \label{figPssAds}}
\end{center}
\end{figure} 
With the profile \eqref{eq:hbads} the temperature \eqref{eq:adstemp1} reads
\begin{equation}
 T=\frac{1}{4\pi r_+}\left[\left(1+\frac{r_+^2}{b^2}\right)\left(1-2\frac{\ellp^2+\frac{4\ellp^4}{b^2-\ellp^2}}{r_+^2+\ellp^2+\frac{4\ellp^4}{b^2-\ellp^2}}\right)+\frac{2r_+^2}{b^2}\right]\,.
 \label{eq:holoadstemp}
  \end{equation}  
The temperature plot is given in Fig. \ref{figPssAds}. At large distances the curve coincides with the conventional Schwarzschild-AdS temperature, while at short distances the black hole cools down in the same way as in Fig. \ref{fig:TCS3d}.  
From the horizon equations one obtains the  internal energy, that reads
\begin{equation}
M =\frac{r_+^2+\ellp^2+\frac{4\ellp^4}{b^2-\ellp^2}}{2r_+\ellp^2}\left(1+\frac{r_+^2}{b^2}\right).
\label{eq:Mbexpl}
\end{equation}
We can also calculate the black hole entropy 
\begin{equation}
S=\int_{\ellp}^{r_{+}}\frac{\d M}{T}= \frac{\pi \left( r_{+}^{2}-\ellp^{2} \right)}{\ellp^{2}}+\frac{2 \pi \left(b^{2}+3\ellp^{2}\right)}{b^{2}-L_{P}^{2}}\ln\left(\frac{r_+}{\ellp}\right). \label{eq:entropyAdS}
\end{equation}
We find a logarithmic correction to the area law. Contrary to the standard Scharzschild-AdS case, the entropy depends on the AdS radius $b$ through the  logarithmic term. If $b\to\ellp$, the entropy diverges, $S\to\infty$. From the point of view of black hole pair creation, this is equivalent to saying that no black hole can form since the AdS-space has attained the maximum entropy configuration. Accordingly the black hole mass parameter \eqref{eq:Mbexpl} vanishes in this limit.

\begin{figure}[t!]
\begin{center}
\includegraphics[width=0.75\textwidth]{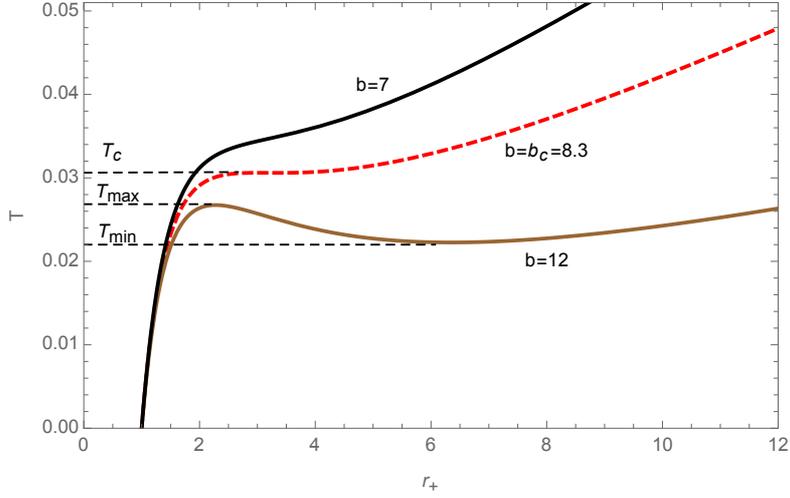}
\caption{Black hole temperature for different values of $b$ vs the temperature of the Schwarzschild-AdS black hole for $\ellp=1$.}
\label{fig:tempholads}
\end{center}
\end{figure} 

 By analyzing the temperature profile in Fig. \ref{fig:tempholads}, one see that there exists a black hole for any temperature. This means that the pure thermal AdS radiation never occurs for $b>\ellp$. Furthermore one can see that there exists a critical value $b_{\rm c}\simeq 8.3 \ellp$ of the AdS radius parameter  such that:
\begin{itemize}
\setlength{\itemsep}{-\parsep}%
\item for $b>b_{\rm c}$, the temperature admits both a minimum value $T_{\rm min}$ and a maximum value $T_{\rm max}$;
\item for $b<b_{\rm c}$, the temperature is a monotonically increasing function of $r_+$;
\item for $b=b_{\rm c}$, the minimum and the maximum temperature coalesce, $T_{\rm min}=T_{\rm max}=T_{\rm c}$, in a point where the curve exhibit an inflection, ${\rm d}T/{\rm d}r_+={\rm d}^2T/{\rm d}r_+^2=0$.
\end{itemize}

\begin{figure}[t!]
\begin{center}
\includegraphics[width=0.75\textwidth]{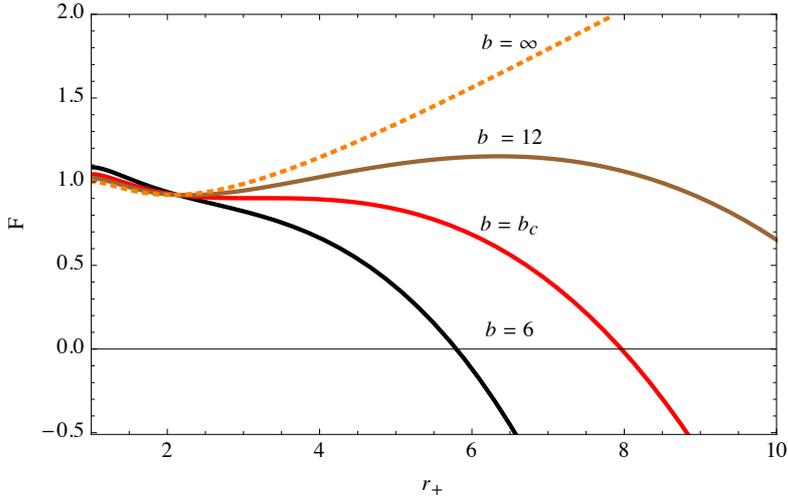}
\caption{Free energy vs horizon radius for $\ellp=1$}
\label{fig:Fvshorizon}
\end{center}
\end{figure} 
The above thermodynamic quantities let us define the free energy of the system
\begin{equation}
F\equiv M-TS.
\end{equation}

\begin{figure}[t!]
\begin{center}
\includegraphics[width=0.75\textwidth]{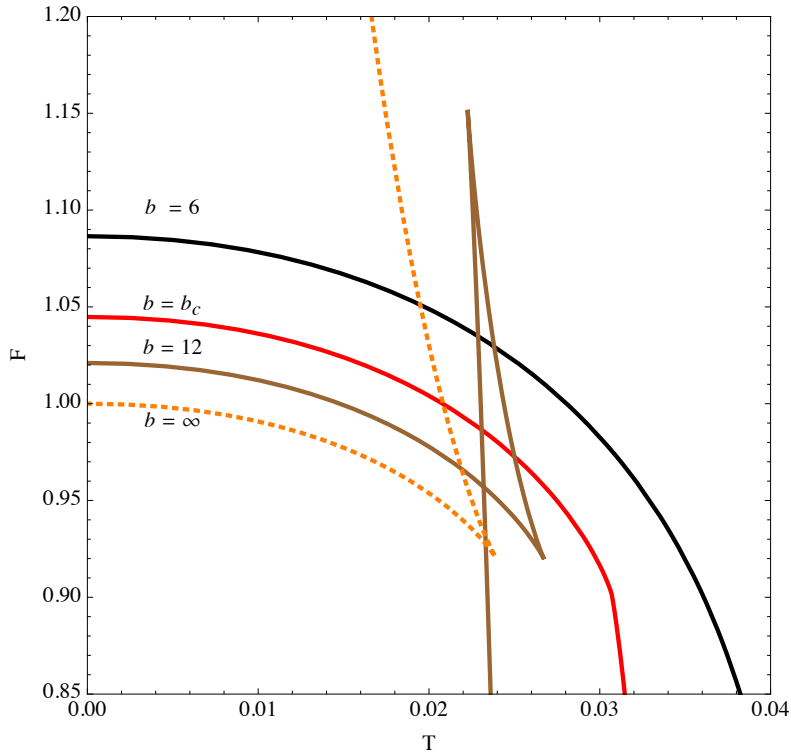}
\caption{Free Energy as function of the temperature for $\ellp=1$. }
\label{eq:FvsTholoads}
\end{center}
\end{figure}

In Fig. \ref{fig:Fvshorizon}, one has the free energy $F$  as a function of the horizon radius $r_+$. We see that above the critical value, $b>b_{\rm c}$, there exists a local minimum of the free energy, while below the critical value $b<b_{\rm c}$, the free energy is monotonically decreasing. In Fig. \ref{eq:FvsTholoads}, there is the free energy as a function of the temperature. For $b>b_{\rm c}$, the curves exhibit a swallow-tail, a typical signature of non-analytic phase transition with latent heat exchange while for $b<b_{\rm c}$ there is a just smooth cross over.

A deeper analysis reveals the following. We start from the case $b>b_{\rm c}$ and we present the following possibilities:
\begin{enumerate}[i)]
\setlength{\itemsep}{-\parsep}%
\item for $T<T_{\rm min}$, there exist just one horizon radius $r_1$. The heat capacity of the black hole $C=\d M/\d T$ is positive defined, the systems is locally stable but the free energy is positive. This means that the black hole is a meta-stable state.  The AdS-thermal background results the favorable state.
\item  For $T_{\rm min}<T<T_{\rm max}$, there are three horizon radii, $r_1<r_2<r_3$. While at $r_1$ and $r_3$ the heat capacity is positive, this is not the case at $r_2$ that is locally unstable. This corresponds the case of a mixed phase. The configuration in $r_2$ can decay to $r_1$ to $r_3$ or in thermal AdS. From Fig. \ref{fig:Fvshorizon}, one can see that $r_1$ is always globally unstable since $F(r_1)>0$, while $r_3$ may have either positive or negative free energy. \item From Fig. \ref{eq:FvsTholoads} we see that there is a value $T_{\rm tr}$ at which the curve intersects itself by forming a cross at the vertex of the swallow tail. The value $T_{\rm tr}$ is the temperature at which the phase transition  between small and big black holes occurs \cite{SpS13}. 
   For $T_{\rm min}< T < T_{\rm tr}$ the pure AdS background is still favorable state, even if small black holes, $r_1$, and big black holes, $r_3$, are possible.  The smaller black hole $r_1$ is the meta-stable state since it sits in local minimum of the free energy in Fig \ref{fig:Fvshorizon}.
\item For $T_{\rm tr}< T< T_{\rm HP}$ the AdS thermal state is still favorable state and the bigger black holes are in the meta-stable state since $F(r_1)>F(r_3)>0$. Here $T_{\rm HP}$ is the Hawking-Page temperature, \textit{i.e.}, $F(T_{\rm HP})=0$.
For $T_{\rm HP}< T <T_{\rm max}$, bigger black holes are the favorable state while AdS radiation and $r_1$ are  meta-stable states. This can be seen by the fact that $F(r_3)<0$. For $T>T_{\rm max}$, there exist only one horizon radius and the black hole is the favorable state while the AdS radiation is the meta-stable state. From Fig. \ref{fig:tempholads}, we see that $T_{\rm max}\simeq 0.03\ellp$.
\item As in the standard Schwarzschild-AdS case there exists a temperature $T_{\rm coll}$, above which the AdS thermal radiation cannot longer sustain itself. The value of such a temperature is of the order $T_{\rm coll}\sim (\ellp/b)^{1/2}\Mpl$ \cite{HaP83}. Accordingly there exist another regime: for $T>T_{\rm coll}$, the AdS radiation will inevitably collapse in a black hole. For $b< b_{\rm coll}$, the collapse of the AdS radiation occurs above the local maximum temperature, \textit{i.e.}, $T_{\rm coll}>T_{\rm max}$.
A rough estimate of parameters suggests that  $b_{\rm coll}\sim 10^3\ellp$.
\end{enumerate}

When $b=b_{\rm c}$ we are at the critical point. This means that the latent heat goes to zero, corresponding to a second order phase transition. Finally, for $b<b_{\rm c}$ one ends up with an analytical cross-over between small and large black holes. In both cases, the bigger black holes are the favorable state. For $T<T_{\rm HP}$, AdS is still the favourable state.

In this section we have seen that the phase structure of the solution \eqref{eq:g00holoads} differs from that of the Schwarzschild-AdS geometry and 
resembles that of models admitting an extremal horizon (see \textit{e.g.} the case of charged black holes {\cite{CEJ99}}, rotating black holes \cite{Caldarelli} and neutral static regular black holes  {\cite{NiT11,Fra16,SmS13}}).  Finally the current results can pave the way to further developments. At the light of recent proposals about the role of the cosmological constant as a thermodynamical variable analogue to the pressure \cite{SmS13,CrM95,Caldarelli,KRT09,Dol11,Dol11b,Dol11c,Dol12,CGK11,LaC12,LaM12,
Gib12,KuM12,GMK12,BCE12,LPP12,HeV13}, it would be interesting to extend the holographic metric \eqref{eq:g00holoads} to the case in which the black hole mass plays the role of the enthalpy of spacetime. We believe this kind of analysis might disclose further insights about the quantum $N$-portrait proposal.

\section{Conclusions}
\label{sec:concl}

In this paper we have offered an extensive analysis of the properties of so called holographic metric \cite{NiS12}. We showed that the new line element describes a neutral static black hole and captures the main features of the quantum $N$-portrait. Specifically the new metric encodes the self-complete character of gravity, \textit{i.e.}, the possibility of masking the curvature singularity with the formation of an extremal horizon at the Planck scale. We also showed that such geometric description  uniquely depends on the parameter $N$, that regulates the pixelization of the event horizon. Interestingly our analysis offers a viable scenario beyond the $N\gg1$ limit usually considered within the quantum $N$-portrait.  We also showed that having a generic $N$, corresponds to replacing the Einstein-Hilbert action with a non-local gravity action. 

We derived the higher-dimensional holographic metric. We showed that the thermodynamic properties are equivalent to those of the four dimensional solution. During the evaporation the black hole reaches a maximum temperature before cooling down to a terascale remnant. The profile of the temperature suggests that the black hole emission is in marked contrast with that of the higher dimensional Schwarzschild black hole. Within the ADD model one can conclude that the holographic black hole tends to emit softer particle mainly on the brane.

Finally we devoted our interest to the holographic metric in the presence of AdS cosmological term. After deriving a new line element, that consistently matches the features of the quantum $N$-portrait with AdS background, we offered a deep analysis of the phase structure of the related thermodynamics. We showed that new metric phase diagram extends the conventional Hawking-Page scenario to a Van der Waals like structure. There exists a critical value of the AdS radius above which the transition from small to large black holes is non-analytical, \textit{i.e.}, with latent heat. In other words, one has a first order transition between small
and large black holes. Below the critical value one has just smooth cross-over, while at the critical value the latent heat is vanishing and one ends up with a second order phase transition.  We believe that translation of the quantum $N$-portrait in a geometric model will lead to the identification of further black hole universal characters beyond the semiclassical limit.

\subsection*{Acknowledgements}
 
 This work has been supported by the project ``Evaporation of the microscopic black holes'' of the German Research Foundation (DFG) under the grant NI 1282/2-1,  by the HIC for FAIR
within the framework of the LOEWE program (Landesoffensive zur Entwicklung Wissenschaftlich -\"{O}konomischer Exzellenz) launched by the State of Hesse, by the H-QM and partially
by the European COST action MP0905 ``Black Holes in a Violent Universe''.\\
The authors thank the referees for the constructive comments.
   


\end{document}